\documentclass[11pt]{iopart}
\usepackage{iopams,amssymb} 
\usepackage{graphicx}
\usepackage{epsfig}
\usepackage[latin1]{inputenc}

\newcommand{\beq}{\begin{equation}}
\newcommand{\eeq}{\end{equation}}
\newcommand{\bqa}{\begin{eqnarray}}
\newcommand{\eqa}{\end{eqnarray}}

\def\lsim{\mathrel{\rlap{\lower4pt\hbox{$\sim$}}
    \raise1pt\hbox{$<$}}}                
\def\gsim{\mathrel{\rlap{\lower4pt\hbox{$\sim$}}
    \raise1pt\hbox{$>$}}}                

\begin{document}

\title{QGP collective effects and jet transport}

\author{Bj\"orn Schenke$^1$, Adrian Dumitru$^1$, Yasushi Nara$^2$\\ and Michael Strickland$^1$}
\address{$^1$Institut f\"ur Theoretische Physik \\
  Johann Wolfgang Goethe - Universit\"at Frankfurt \\
  Max-von-Laue-Stra\ss{}e~1,
  D-60438 Frankfurt am Main, Germany\vspace*{2mm}}
\address{$^2$Akita International University 193-2 Okutsubakidai \\
  Yuwa-Tsubakigawa, Akita-City, Akita 010-1211, Japan \\
  \vspace*{2mm}}

\begin{abstract}
We present numerical simulations of the SU(2) Boltzmann-Vlasov equation
including both hard elastic particle collisions and soft interactions
mediated by classical Yang-Mills fields. We provide an estimate of the
coupling of jets to a hot isotropic plasma which is independent of infrared cutoffs. 
In addition, we investigate jet propagation in anisotropic plasmas, as created in
heavy-ion collisions.
The broadening of jets is found to be stronger along the beam line than in azimuth,
due to the creation of field configurations with  $B_\perp>E_\perp$ and $E_z>B_z$
via plasma instabilities.

\end{abstract}
\pacs{12.38.-t, 12.38.Mh, 24.85.+p}

\section{Introduction}
High transverse momentum jets produced in heavy-ion collisions
represent a valuable tool for studies of the properties of the hot
parton plasma produced in the central rapidity
region~\cite{Jacobs:2005pk}. 
However, present estimates of the strength of the coupling 
of jets to a QCD plasma are sensitive to infrared cutoffs.
We employ a numerical simulation of the Boltzmann-Vlasov equation, which is 
coupled to the Yang-Mills equation for the soft gluon degrees of freedom.
Soft momentum exchanges between particles are mediated by the fields, while
hard momentum exchanges are described by a collision term including
binary elastic collisions.
This way, we are able to provide an estimate of the coupling of jets to a hot 
plasma which is independent of infrared cutoffs.

The longitudinal expansion of the plasma may lead to a strongly
anisotropic momentum distribution in the local rest frame,
during the very early stages of the plasma evolution.
Due to anisotropies in the particle momentum distributions
plasma instabilities appear~\cite{Romatschke:2003ms}.
These lead to the formation of long-wavelength
chromo-fields with $E_z>B_z$ and $B_\perp >E_\perp$,
which affect the propagation of a hard jet and of its induced hard radiation field.
This may provide an explanation for
the observed asymmetry in measurements of dihadron-correlations.
Here a much stronger broadening of jets in pseudorapidity ($\eta$) than in azimuthal
angle ($\phi$) has been observed~\cite{Jacobs:2005pk,Putschke:2007mi}.

\section{Boltzmann-Vlasov equation for non-Abelian gauge theories}
We solve the classical transport equation for hard gluons with SU(2)
color charge 
including hard binary collisions
\begin{eqnarray}
    p^\mu \left[\partial_\mu + g q^a F_{\mu\nu}^a \partial^\nu_{p} 
    + g f^{abc} A_\mu^b(x) q^c \partial_{q^a} \right]f={\cal C}\,,
\end{eqnarray}
where $f=f(x,p,q)$ denotes the single-particle phase space distribution.
It is coupled self-consistently to the Yang-Mills equation for the
soft gluon fields.
%
The collision term contains all binary collisions, descibed 
by the leading-order $gg\rightarrow gg$ tree-level diagrams.

We replace the distribution $f(x,p,q)$ by a large number of test
particles, which leads to Wong's equations~\cite{Wong:1970fu}
\begin{eqnarray}
    \dot{\mathbf{x}}_i(t)&=\mathbf{v}_i(t)\,,~~~\dot{\mathbf{p}}_i(t)=g
    q^a_i(t)\left(\mathbf{E}^a(t)+\mathbf{v}_i(t)\times\mathbf{B}^a(t)\right)
    \,,\nonumber \\
    \dot{q}_i(t)&=-igv_i^\mu(t)[A_\mu(t),q_i(t)]\,,\label{wong}
\end{eqnarray}
for the $i$-th test particle, whose coordinates are $\mathbf{x}_i(t)$,
$\mathbf{p}_i(t)$, and $q^a_i(t)$. The time evolution of the
Yang-Mills field is determined by the standard Hamiltonian method
\cite{Ambjorn:1990pu} in $A^0=0$ gauge. See~\cite{HuBM,Dumitru:2005gp,Dumitru:2006pz,Dumitru:2007rp}
for more details.
The collision term is incorporated using the stochastic method
\cite{Danielewicz:1991dh}. 
The total cross section is given by
$
    \sigma_{2\to2}=\int_{k^{*2}}^{{s}/{2}}\frac{d\sigma}{dq^2}dq^2\,,
$
where we have introduced a lower cutoff $k^*$.
To avoid double-counting, this cutoff
should be on the order of the hardest field mode that can be
represented on the given lattice, $k^*\simeq\pi/a$, with the lattice spacing $a$. 
For a more detailed discussion see \cite{Dumitru:2007rp}.


\section{Jet broadening in an isotropic plasma}
\begin{figure}[t]
  \hfill
  \begin{minipage}[t]{.48\textwidth}
  \begin{center}
    \includegraphics[width=6.5cm]{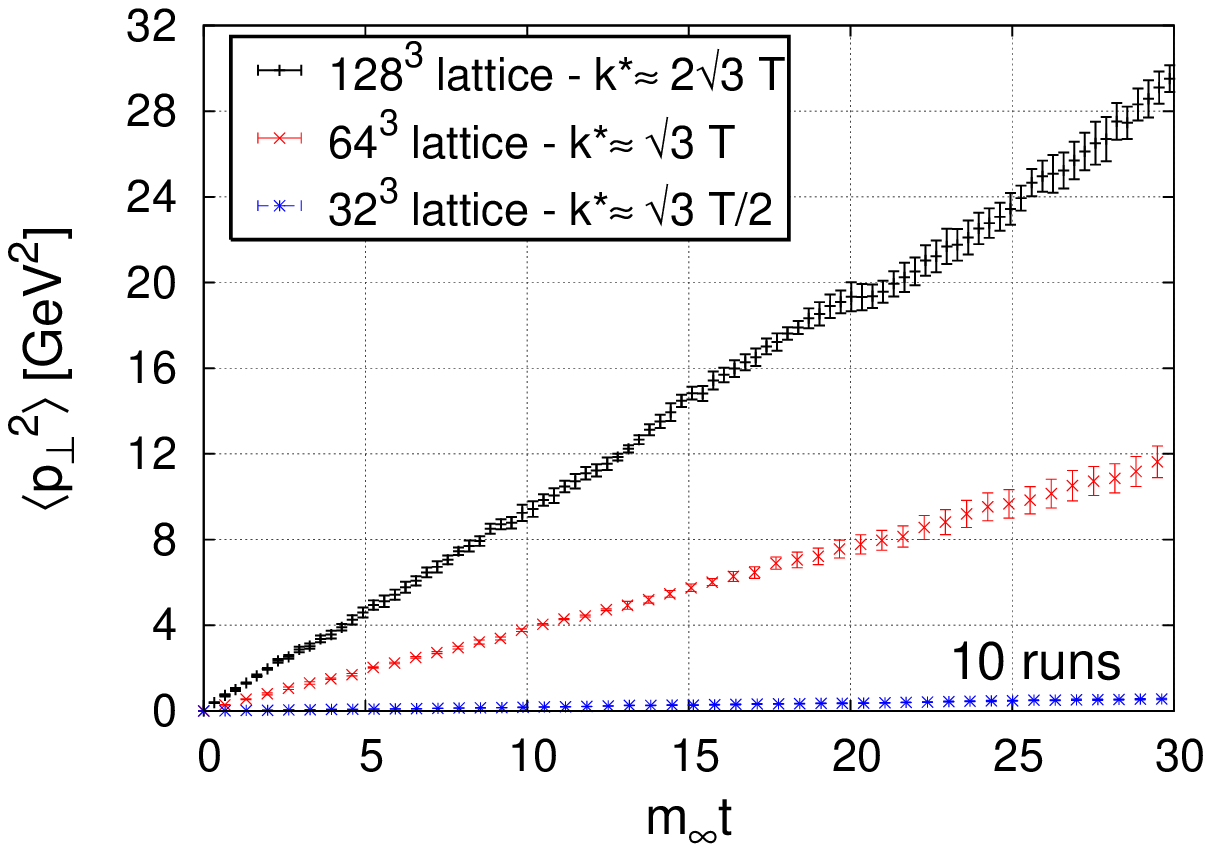}
    \caption{Momentum diffusion caused by
    particle-field interactions only.}
    \label{fig:ptnocoll}
  \end{center}
  \end{minipage}
  \hfill
  \begin{minipage}[t]{.48\textwidth}
  \begin{center}
    \includegraphics[width=6.5cm]{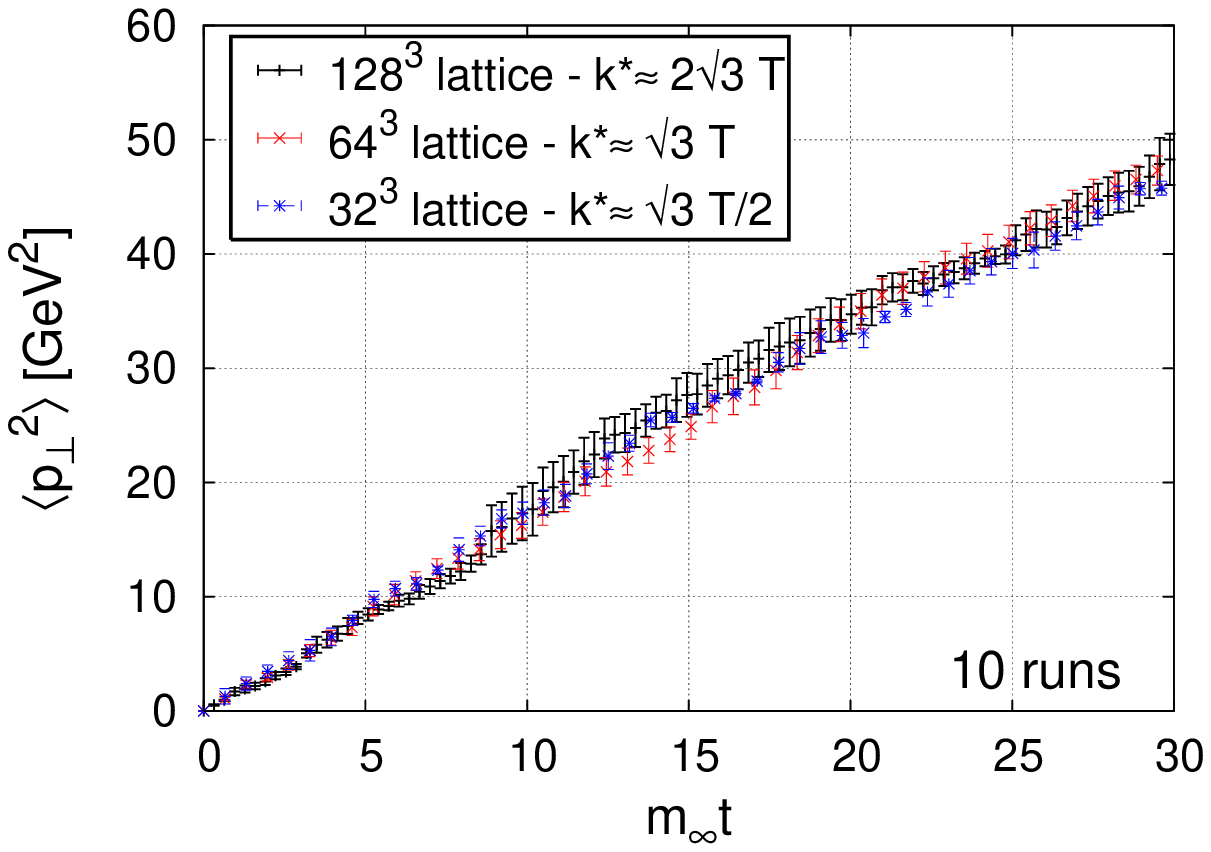}
    \caption{Momentum diffusion by both particle-field
    and direct particle-particle interactions.}
    \label{fig:ptcoll}
  \end{center}
  \end{minipage}
  \hfill
\end{figure}

We first consider a heat-bath of particles with a density of
$n_g=10/{\rm fm}^3$ and an average particle momentum of
$3T=12$~GeV. 
For a given lattice (resp.\ $k^*$) we take the initial
energy density of the thermalized fields to be $\int d^3k/(2\pi)^3 \,
k \hat{f}_{\rm Bose}(k)\Theta(k^*-k)$, where $\hat{f}_{\rm Bose}(k)=n_g/(2T^3 \zeta(3))/(e^{k/T}-1)$ is a Bose
distribution normalized to the assumed particle density $n_g$, and $\zeta$ is the Riemann zeta function.
The initial spectrum is fixed to Coulomb gauge and $A_i\sim 1/k$.
We measure the momentum broadening $\langle p_\perp^2\rangle(t)$
of high-energy test particles ($p/3T\approx 5$) passing through this
medium. Fig.~\ref{fig:ptnocoll} shows that in the collisionless case,
${\cal C}=0$, the broadening is stronger on larger lattices which
accommodate harder field modes. However, Fig.~\ref{fig:ptcoll}
demonstrates that collisions with momentum exchange larger than
$k^*(a)$ compensate for this growth and lead to approximately
lattice-spacing independent results.


A related transport coefficient is $\hat
q$~\cite{Baier:1996sk}. It is the typical momentum transfer (squared)
per collision divided by the mean-free path, which is nothing but
$\langle p_\perp^2\rangle(t)/t$. From Fig.~\ref{fig:ptcoll}, $\hat
q\simeq 2.2$~GeV$^2$/fm for $N_c=2$, $n_g=10/$fm$^3$ and
$p/(3T)\approx5$. 
We have verified that $\hat q$ does not depend on
the temperature $T$ as long as the particle density $n_g$ and the
ratio $p/T$ is fixed.
Due to the independence of $\hat{q}$ of the temperature and its proportionality to the density $n$,
we can scale to physical densities for a QGP created at RHIC.
We adjust for the different color factors in SU(3), and find $\hat{q} \approx 5.6$
GeV$^2/$fm, at $T=400$ MeV, $E_{{\rm jet}}\approx 20$ GeV ($p/3T
= 16$) in a system of quarks and gluons.


\section{Jet broadening in an unstable plasma}
\begin{figure}[htb]
  \begin{center}
    \includegraphics[width=14cm]{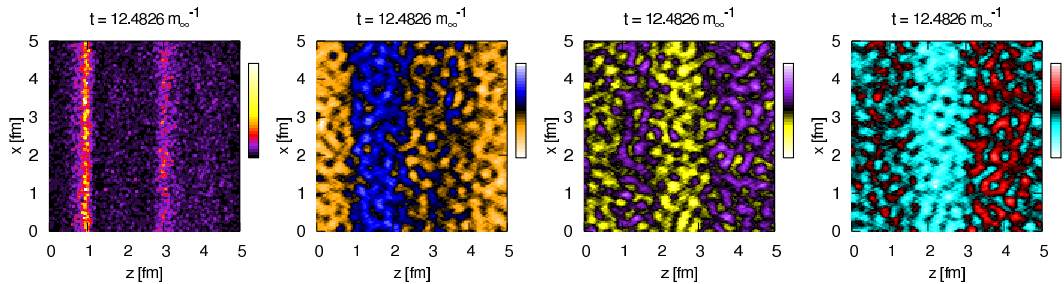}
    \caption{Slices in the $x$-$z$-plane at fixed $y=L/2$ of the current in the $x$-direction, 
$J_x$, and the three color components of the chromo-magnetic field in the
$y$-direction. Filaments are nicely 
visible. Scales in lattice units. $0$ to $5 \cdot 10^{-8}$ for the
current, $-4 \cdot 10^{-3}$ to $4 \cdot 10^{-3}$ for the chromo-magnetic fields.}
    \label{fig:thns}
  \end{center}
\end{figure}

In heavy-ion collisions, locally anisotropic momentum distributions
may emerge due to the longitudinal expansion. Such anisotropies
generically give rise to instabilities~\cite{Romatschke:2003ms,Dumitru:2005gp,Dumitru:2006pz}.
Here, we investigate their effect on the momentum broadening of jets,
including the effect of collisions.  
The initial anisotropic momentum distribution
for the hard plasma gluons is taken to be
$
f(\mathbf{p})=n_g \left(\frac{2\pi}{p_{{\rm h}}}\right)^2 \delta(p_z)
  \exp(-p_\perp/p_{{\rm h}})\,, 
$
with $p_\perp=\sqrt{p_x^2+p_y^2}$. 
We initialize small-amplitude fields sampled from a
Gaussian distribution and set $k^*\approx p_{{\rm h}}$, for the
reasons alluded to above.
We add additional high momentum particles with $p_x=12\,
p_{{\rm h}}$ and $p_x=6\, p_{{\rm h}}$, respectively, to investigate
the broadening in the $y$ and $z$ directions via the variances
$
	\kappa_\perp(p_x):=\frac{d}{dt}\langle(\Delta
        p_\perp)^2\rangle\,, \kappa_z(p_x):=\frac{d}{dt}\langle(\Delta p_z)^2\rangle\,.
$
The ratio $\kappa_z/\kappa_\perp$ can be roughly associated with the
ratio of jet correlation widths in azimuth and rapidity:
${\kappa_z}/{\kappa_\perp} \approx {\langle\Delta\eta\rangle} /
{\langle\Delta\phi\rangle}$. Experimental data on dihadron
correlation functions for central Au+Au collisions at
$\sqrt{s}=200$~GeV~\cite{Jacobs:2005pk} are consistent with
$\kappa_z/\kappa_\perp\approx 3$~\cite{Romatschke:2006bb}.
\begin{figure}[t]
  \hfill
  \begin{minipage}[t]{.48\textwidth}
  \begin{center}
    \includegraphics[width=6.5cm]{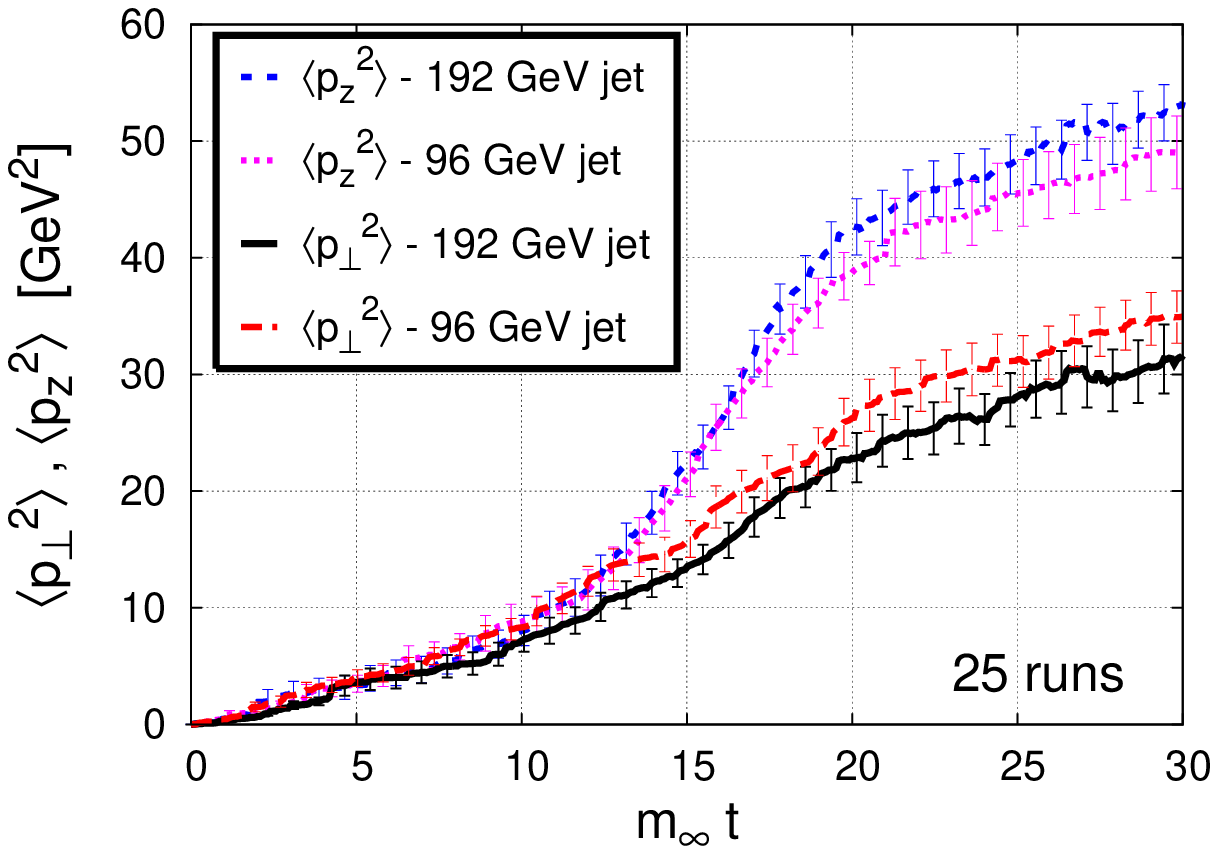}
    \caption{Momentum broadening of a jet transverse to its initial momentum.}
    \label{fig:pxyjet192}
  \end{center}
  \end{minipage}
  \hfill
  \begin{minipage}[t]{.48\textwidth}
  \begin{center}
    \includegraphics[width=7cm]{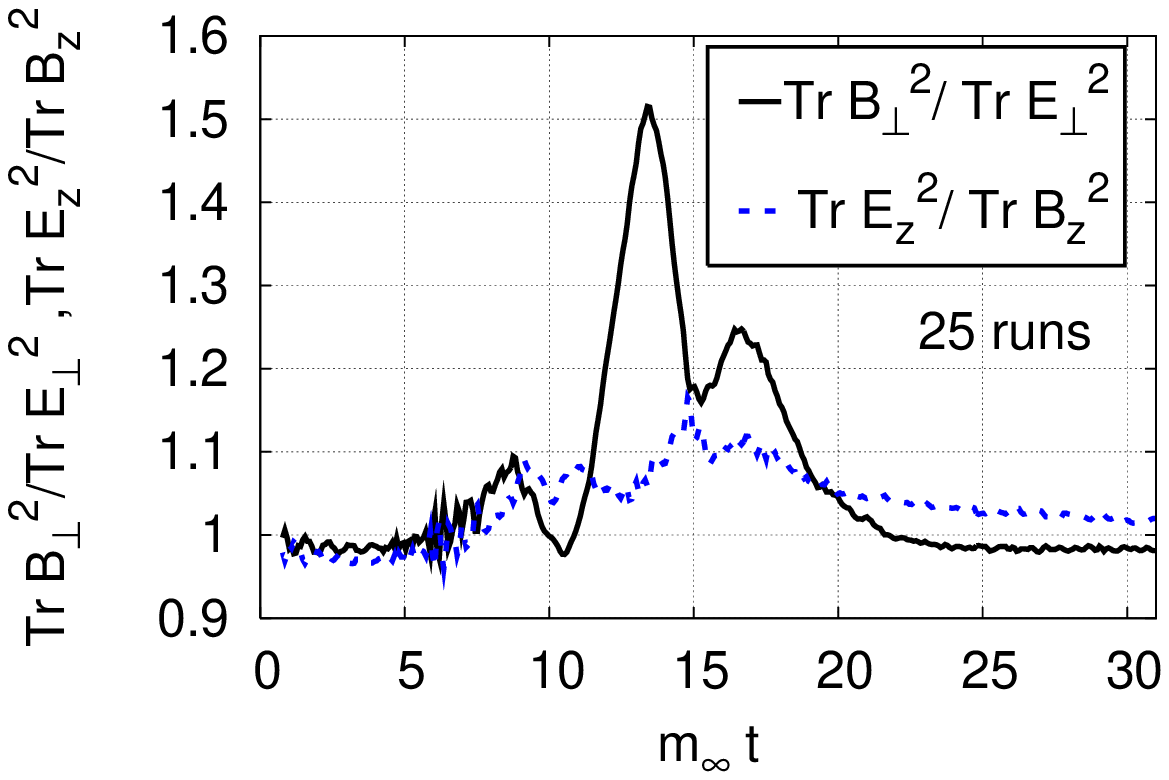}
    \caption{Ratios of field energy densities.}
    \label{fig:ratios}
  \end{center}
  \end{minipage}
  \hfill
\end{figure}
Fig.~\ref{fig:pxyjet192} shows the time evolution of $\langle
p_\perp^2 \rangle$ and of $\langle p_z^2 \rangle$.
During the period of instability and for both jet energies we find
$
	\kappa_z/\kappa_\perp \approx 2.3\,.
$
The explanation for the larger broadening along the beam axis is as
follows. In the Abelian case the instability generates predominantly
transverse magnetic fields which deflect the particles in
the $z$-direction~\cite{Majumder:2006wi}.
Although the interactions are a lot less trivial, in a non-Abelian plasma the instability 
creates large domains of strong chromo-electric and -magnetic fields with
$E_z>B_z$, aside from $B_\perp>E_\perp$ (Fig.~\ref{fig:ratios}).  
The field configurations are such that particles are deflected preferentially in
the longitudinal $z$-direction (to restore isotropy).
Fig. \ref{fig:thns} shows the filamentation of the current and the domains of 
magnetic fields generated by the instability.
~\\\emph{Acknowledgments:~} M.S.\ and B.S.\ are supported by DFG Grant GR 1536/6-1.
~\\


\end{document}